\documentclass{ws-procs9x6}

\begin{document}
\title{Symmetry in Cartan language\\for geometric theories of gravity}

\author{M. Hohmann$^*$}

\address{Laboratory of Theoretical Physics, Institute of Physics, University of Tartu,\\
Tartu, Tartumaa, Estonia\\
$^*$E-mail: manuel.hohmann@ut.ee\\
http://kodu.ut.ee/\textasciitilde{}manuel/}

\begin{abstract}
We present a recent definition of symmetry generating vector fields on manifolds equipped with a first-order reductive Cartan geometry. We apply this definition to a number of spacetime geometries used in gravity theories and show that it agrees with the usual notions of symmetry of affine, Riemann-Cartan, Riemannian, Weizenb\"ock and Finsler spacetimes.
\end{abstract}


\bodymatter

\section{Definition}\label{sec:definition}
Let \(M\) be a manifold and \(\varphi: \mathbb{R} \times M \to M\) a one-parameter group of diffeomorphisms generated by a vector field \(\xi\) on \(M\). On the general linear frame bundle
\begin{equation}\label{eqn:glframebundle}
\mathrm{GL}(M) = \bigcup_{x \in M}\{\text{linear bijections } f: \mathbb{R}^n \to T_xM\}
\end{equation}
we define a one-parameter group of diffeomorphisms \(\bar{\varphi}: \mathbb{R} \times \mathrm{GL}(M) \to \mathrm{GL}(M)\) by \(\bar{\varphi}_t(f) = \varphi_{t*} \circ f\). This one-parameter group is generated by a vector field \(\bar{\xi}\) on \(\mathrm{GL}(M)\), which we call the frame bundle lift of \(\xi\).

Let \(G\) be a Lie group with closed subgroup \(H \subset G\), \(\pi: P \to M\) a principal $H$-bundle with \(P \subset \mathrm{GL}(M)\) and \(A \in \Omega^1(P,\mathfrak{g})\) a Cartan connection which is first order reductive, i.e., the adjoint representation of \(H\) on the Lie algebra \(\mathfrak{g}\) splits into subrepresentations \(\mathfrak{g} = \mathfrak{h} \oplus \mathfrak{z}\) and the resulting representation on \(\mathfrak{z}\) is faithful, and such that the $\mathfrak{z}$-valued part \(e \in \Omega^1(P,\mathfrak{z})\) of \(A\) is the solder form on \(P\). We call the Cartan geometry \((\pi: P \to M, A)\) \emph{invariant}\cite{Hohmann:2015pva} under a vector field \(\xi\) on \(M\) if and only if the frame bundle lift \(\bar{\xi}\) is tangent to \(P\) and its restriction to \(P\) preserves \(A\), i.e., \(\mathcal{L}_{\bar{\xi}}A = 0\).

\section{Applications}\label{sec:applications}
We apply the notion of invariance defined above to a number of common spacetime geometries used in various models of gravity. These geometries can be written as first-order reductive Cartan geometries. In particular, we obtain the following notions of invariance under a vector field \(\xi\):
\begin{itemlist}
\item
Affine geometry with connection \(\Gamma\):
\begin{equation}
\mathcal{L}_{\xi}\Gamma = 0\,.
\end{equation}
\item
Riemann-Cartan geometry with metric \(g\) and torsion \(T\):
\begin{equation}
\mathcal{L}_{\xi}g = 0 \quad \wedge \quad \mathcal{L}_{\xi}T = 0\,.
\end{equation}
\item
Riemannian geometry with metric \(g\):
\begin{equation}
\mathcal{L}_{\xi}g = 0\,.
\end{equation}
\item
Weizenb\"ock geometry with tetrad \(e\):
\begin{equation}
\mathcal{L}_{\xi}e = \lambda e\,,
\end{equation}
where \(\lambda\) is a constant infinitesimal Lorentz transformation.
\item
Finsler geometry\cite{Hohmann:2013fca} with Finsler length function \(F\):
\begin{equation}
\mathcal{L}_{\hat{\xi}}F = 0\,,
\end{equation}
where \(\hat{\xi}\) is the tangent bundle lift of \(\xi\), i.e., the vector field on \(TM\) which is defined in analogy to the frame bundle lift \(\bar{\xi}\), but replacing the diffeomorphisms \(\bar{\varphi}_t\) with \(\hat{\varphi}_t = \varphi_{t*}\).
\end{itemlist}
These notions of invariance agree with the standard notions of invariance on the respective spacetime models.

\section*{Acknowledgments}
The author gratefully acknowledges the full financial support of the Estonian Research Council through the Postdoctoral Research Grant ERMOS115 and the Startup Research Grant PUT790.

\end{document}